\documentclass[letterpaper]{article} % DO NOT CHANGE THIS
\usepackage{aaai2027}  % DO NOT CHANGE THIS
\usepackage[hyphens]{url}  % DO NOT CHANGE THIS
\usepackage{graphicx} % DO NOT CHANGE THIS
\usepackage{natbib}  % DO NOT CHANGE THIS AND DO NOT ADD ANY OPTIONS TO IT
\usepackage{caption} % DO NOT CHANGE THIS AND DO NOT ADD ANY OPTIONS TO IT
\usepackage{algorithm}
\usepackage{algorithmic}
\usepackage{amsmath} 
\usepackage{makecell}
\usepackage{tabularx}
\usepackage{array}
\usepackage{multirow}

\newcolumntype{Y}{>{\raggedright\arraybackslash}X}
\usepackage{newfloat}
\usepackage{listings}
\DeclareCaptionStyle{ruled}{labelfont=normalfont,labelsep=colon,strut=off} % DO NOT CHANGE THIS
\floatstyle{ruled}
\newfloat{listing}{tb}{lst}{}
\floatname{listing}{Listing}

\usepackage{booktabs}

\nocopyright
\title{CommBench: Can LLMs Write Correct and Efficient GPU Communication Code?}
\author{
    Shuang Ma\textsuperscript{\rm 1},
    Yuyi Li\textsuperscript{\rm 1},
    Yihan Zhang\textsuperscript{\rm 1},
    Hezhi Xie\textsuperscript{\rm 1},
    Danyang Chen\textsuperscript{\rm 4},\\
    Shuyang Ji\textsuperscript{\rm 3},
    Ziming Mao\textsuperscript{\rm 2},
    Cheng Ji\textsuperscript{\rm 3},
    Ansha Prashanth\textsuperscript{\rm 1},
    Wenting Yang\textsuperscript{\rm 5},\\
    Yiran Wang\textsuperscript{\rm 3},
    Chihan Cui\textsuperscript{\rm 6},
    Pei Yu Lin\textsuperscript{\rm 1},
    Ion Stoica\textsuperscript{\rm 2},
    Yang Zhou\textsuperscript{\rm 1}
}

\affiliations{
    \textsuperscript{\rm 1}University of California, Davis\\
    \textsuperscript{\rm 2}University of California, Berkeley\\
     \textsuperscript{\rm 3} University of Illinois at Urbana-Champaign \\
     \textsuperscript{\rm 4} The Chinese University of Hong Kong\\
     \textsuperscript{\rm 5}University of California, San Diego\\
     \textsuperscript{\rm 6}University of Wisconsin--Madison\\
}

\begin{document}

\maketitle

\begin{abstract}
Training and serving large language models (LLMs) rely heavily on high-performance GPU communication, yet implementing efficient GPU communication primitives requires deep expertise in GPU architectures, networking hardware, and distributed communication patterns, making them particularly challenging for code generation models. We present CommBench, a comprehensive benchmark for GPU communication programming, consisting of over 100 expert-curated tasks spanning point-to-point communication, collective operations, expert-parallel communication, compute--communication fusion, and communication utility functions, with reference implementations either written by GPU communication experts or distilled from production codebases. We further introduce a cheat-resistant evaluation framework that automatically compiles, executes, and validates generated code on multi-GPU systems, and a unified metric that jointly measures functional correctness and communication performance. Evaluating leading frontier and open-source code generation models on both intra-node NVLink and inter-node RDMA platforms reveals that even the strongest model, GPT-5.5, correctly implements and achieves competitive performance on only 30.7\% of the benchmark tasks. Our results expose a substantial gap between current LLMs and expert-written GPU communication code, establishing CommBench as a challenging benchmark for advancing AI-assisted systems programming.

\end{abstract}


\section{Introduction}
GPU communication, including both standalone communication primitives and communication-computation fusion, has become a cornerstone of modern large language model (LLM) training and inference. As LLMs continue to scale in model size, context length, and deployment, distributed execution increasingly relies on efficient inter-GPU communication for tensor, pipeline, expert, and context parallelism, distributed optimization, and KV-cache transfer \cite{shoeybi2020megatronlmtrainingmultibillionparameter, narayanan2021efficient, deepseekv3, kwon2023vllm, MLSYS2026_e01c431b}. Consequently, communication has emerged as a major performance
bottleneck: it can account for 43.6\% of forward-pass time when
training a production Mixture-of-Experts (MoE) model and up to 47\% of end-to-end model
execution time in distributed MoE workloads
\cite{jin2026megascale,zhang2025comet}.
 %% verify the number
% GPU communication, including fused communication .. is important. xxx. 

Driven by the rapid evolution of LLM systems and GPU architectures, the demand for high-performance GPU communication programming continues to grow. Although widely adopted libraries such as NCCL provide efficient implementations of common communication primitives, they are designed as general-purpose abstractions and cannot always capture workload-specific optimizations required by modern LLM systems \cite{nccl, hwang2026mscclpp}. Consequently, many industrial LLM frameworks maintain customized communication stacks to support architecture-aware optimizations, communication-computation fusion, and emerging execution paradigms \cite{deepseekv3, deepep, mooncake}. Meanwhile, modern accelerators such as NVIDIA Hopper, Blackwell, and AMD Instinct GPUs continue to increase computational throughput, making communication efficiency increasingly critical for sustaining hardware utilization. This trend has motivated GPU-initiated communication mechanisms that reduce CPU involvement and enable lower-latency execution \cite{gdasync, hamidouche2025gin}. Furthermore, emerging LLM architectures, such as MoE, introduce highly irregular and fine-grained communication patterns that are poorly served by traditional collective communication libraries, driving the development of specialized communication systems and kernels \cite{deepseekv3, jin2026megascale, zhang2025comet}.

GPU communication programming is substantially more challenging than single-GPU programming for three key reasons. First, it requires specialized expertise spanning GPU kernel optimization, distributed systems, networking, and accelerator architectures \cite{cuda, nccl, hwang2026mscclpp, hamidouche2025gin}. Second, developers must coordinate computation and communication across multiple devices over complex and failure-prone interconnects while ensuring both correctness and high performance, making implementation and debugging significantly more difficult than single-device programs \cite{hoefler2015correctness, deepep, jin2026megascale, zhang2025comet}. Finally, unlike conventional code generation benchmarks, GPU communication lacks large-scale, realistic, and faithfully executable datasets, limiting both the development and evaluation of coding models in this domain \cite{jimenez2024swebench, jain2025livecodebench, zhuo2025bigcodebench}.

Despite its growing importance, GPU communication programming remains largely unexplored by existing LLM coding benchmarks. Widely adopted benchmarks, including KernelBench, ComputeEval, and TritonBench, primarily evaluate single-GPU code generation tasks \cite{ouyang2025kernelbenchllmswriteefficient, computeeval, li2025tritonbenchbenchmarkinglargelanguage}. None of them assesses whether LLMs can generate correct and efficient GPU communication programs, which require coordinating multiple GPUs, orchestrating communication over heterogeneous interconnects, and tightly coupling communication with computation. Consequently, existing benchmarks fail to evaluate emerging GPU communication tasks such as implementing communication primitives (e.g., MSCCL++ channels and collective interfaces), GPU-initiated networking, and communication-computation fusion kernels (e.g., fused AllGather--GEMM pipelines spanning NVLink and InfiniBand).

%% TODO: add more benchmark related to cuda

% Despite all this, multi-device GPU programming has been largely overlooked in LLM coding benchmarks.

To bridge this gap, we present \textbf{CommBench}, the first comprehensive benchmark for LLM-generated GPU communication code across diverse interconnects, programming abstractions, and GPU vendors. CommBench comprises over 100 GPU communication problems with reference implementations, covering representative industry-scale multi-GPU communication workloads derived from real-world development. It spans point-to-point communication, collective operations, expert-parallel communication, communication--computation fusion, and utility functions. The benchmark is constructed from expert-written implementations and carefully distilled production code from widely used GPU communication frameworks, including MSCCL++, NCCL, NVSHMEM, DeepEP, ThunderKittens, vLLM, and SGLang~\cite{hwang2026mscclpp,nccl,nvshmem,deepep,thunderkittens,kwon2023vllm,zheng2024sglang}. We evaluate leading frontier and open-source code generation models using a cheat-resistant evaluation harness on real hardware spanning both intra-node NVLink and inter-node RDMA environments~\cite{nvlink,rdma-core}, and provide detailed analyses of their strengths and failure modes. Our results reveal that even the strongest model, GPT-5.5, correctly solves and achieves competitive performance on only 30.7\% of CommBench tasks.

% A breif results. 

\section{CommBench and Framework Structure}
\subsection{Benchmark Structure}

CommBench consists of a collection of independently executable examples, currently comprising more than 100 GPU communication tasks. Table~\ref{tab:composition} summarizes the composition of the benchmark.  Each example includes a problem specification and a reference implementation. Some examples implement complete communication functionalities, such as point-to-point and collective communication interfaces or MoE expert-parallel dispatch and combine kernels, while others provide reusable communication primitives, such as MSCCL++ communication channels. Based on our experience developing high-performance GPU communication libraries, we manually categorize each example into one of three difficulty levels: \textit{Easy}, \textit{Medium}, and \textit{Hard}.

The benchmark covers five representative categories of GPU communication programming: (1) \textbf{Point-to-Point (P2P)}, which implements communication between pairs of devices; (2) \textbf{Collective}, including operations such as AllReduce, AllGather, ReduceScatter, Broadcast, and All-to-All; (3) \textbf{Expert Parallel (EP)}, which implements sparse, non-uniform communication patterns for MoE training and inference; (4) \textbf{Communication--Computation Fusion}, which overlaps communication with computation (e.g., fused AllGather--GEMM kernels); and (5) \textbf{Utilities}, including supporting components such as connection establishment, memory registration, synchronization, and topology discovery.

To ensure both realism and diversity, CommBench combines expert-written implementations with production code distilled from state-of-the-art GPU communication libraries and LLM serving frameworks, including CUDA Runtime, libibverbs, MSCCL++, NCCL, NVSHMEM, DeepEP, NCCL Device API, ThunderKittens, vLLM, and SGLang.

% Table~\ref{tab:composition} summarizes the composition of the benchmark. Roughly one third of the examples are \emph{bare-metal}, implemented directly against the CUDA Runtime or libibverbs without any communication library, while the remaining two thirds are \emph{library-based} and exercise the interfaces of a specific GPU communication framework or serving system. No single category, difficulty level, or library dominates the benchmark, and the long tail of specialized libraries is deliberate: it is precisely the code that is scarce in web-scale pre-training corpora.

\begin{table}[t]
\centering
\small
\setlength{\tabcolsep}{3.5pt}
\caption{Composition of CommBench ($N=101$ examples). Each panel partitions the full benchmark; percentages are computed over all examples and may not sum to exactly 100\% due to rounding.}
\label{tab:composition}
\begin{tabular}{@{}lrr@{\hskip 10pt}lrr@{}}
\toprule
\multicolumn{3}{c}{\textbf{Task category}} & \multicolumn{3}{c}{\textbf{Library}} \\
\cmidrule(r){1-3}\cmidrule(l){4-6}
 & \# & \% & & \# & \% \\
\midrule
Collective       & 32 & 31.7 & ThunderKittens  & 24 & 23.8 \\
Utilities        & 25 & 24.8 & libibverbs      & 17 & 16.8 \\
Fusion           & 19 & 18.8 & CUDA Runtime    & 14 & 13.9 \\
P2P              & 16 & 15.8 & MSCCL++         & 12 & 11.9 \\
EP               &  9 &  8.9 & vLLM            &  9 &  8.9 \\
\cmidrule(r){1-3}
\multicolumn{3}{@{}l}{\textbf{Difficulty}}   & DeepEP          &  8 &  7.9 \\
\cmidrule(r){1-3}
Easy             & 28 & 27.7 & NCCL Device API &  5 &  5.0 \\
Medium           & 52 & 51.5 & NCCL            &  4 &  4.0 \\
Hard             & 21 & 20.8 & NVSHMEM         &  4 &  4.0 \\
\cmidrule(r){1-3}
\multicolumn{3}{@{}l}{\textbf{Setup}}        & SGLang          &  4 &  4.0 \\
\cmidrule(r){1-3}
Bare-metal       & 31 & 30.7 &                 &    &      \\
Library-based    & 70 & 69.3 &                 &    &      \\
\bottomrule
\end{tabular}
\end{table}

%% TODO: add a statistic graph.
\subsection{Evaluation Framework}

We develop an automated evaluation framework that executes generated programs on real hardware and supports iterative refinement through execution feedback. The framework is carefully designed to ensure faithful evaluation by preventing models from modifying the evaluation environment or circumventing the intended verification procedure.

\paragraph{Example Structure.} Each benchmark example consists of three components.

First, each example includes a \textbf{problem specification} containing the task description together with a partially completed source file, where the implementation is replaced with \texttt{// TODO} while preserving the original function signatures and interfaces. Models are required to complete only the designated implementation regions. During evaluation, we verify that no code outside these editable regions has been modified, preventing models from bypassing the intended task.

Second, a \textbf{reference implementation} provides a high-quality, expert-written solution implemented in Python, CUDA, HIP, or C++. The reference implementation is accompanied by a randomized test harness that validates both correctness and performance. To avoid benchmark contamination, reference implementations are excluded from the public benchmark release and are never exposed to evaluated models.

Finally, a hidden \textbf{build-and-execution script} compiles and executes each solution under a controlled environment. Because the compilation process is entirely managed by the evaluation framework, generated programs are restricted to the intended software stack and cannot introduce unauthorized dependencies.

\paragraph{Iterative Refinement.}
The framework optionally supports multi-round code refinement. If the user-configured maximum refinement round is greater than one, compilation errors, runtime failures, or performance statistics from the previous execution are incorporated into the next prompt as execution feedback. 
% The model iteratively revises its solution until either a correct implementation with improved performance is obtained or the predefined refinement budget is exhausted.

\begin{table*}[t]
\centering
\small
\caption{Overall results on CommBench. Models are ranked by the proposed \emph{Quality-Weighted Pass Rate}, which jointly measures correctness coverage and generated code quality. Higher is better for all metrics except Price.}
\label{tab:leaderboard}
\begin{tabular}{lcccccc}
\toprule
Model &
Quality-Weighted &
Pass &
PASS+Good &
GM-Speedup &
Open &
Price \\
&
Pass Rate $\uparrow$ &
Rate (\%) $\uparrow$ &
(\%) $\uparrow$ &
$\uparrow$ &
Source &
(\$) $\downarrow$ \\
\midrule
\textbf{GPT-5.5}                 & \textbf{0.467} & \textbf{57.4} & \textbf{30.7} & 0.813 & No  & 1.91 \\
Gemini-3.1-Pro-Preview           & 0.305 & 36.6 & 25.7 & 0.832 & No  & 0.26 \\
Claude Opus 4.7                  & 0.282 & 33.7 & 20.8 & 0.836 & No  & 0.21 \\
GLM-5.1                          & 0.281 & 29.7 & 17.8 & 0.947 & Yes & 0.63 \\
Kimi-K2.6                        & 0.275 & 30.7 & 18.8 & 0.895 & Yes & 0.10 \\
Qwen3.7-Max                      & 0.269 & 26.7 & 15.8 & 1.008 & No & 0.03 \\
DeepSeek-V4-Pro                  & 0.197 & 19.8 & 12.9 & 0.995 & Yes & \textbf{0.02} \\
\bottomrule
\end{tabular}
\end{table*}

\section{Experiments}

\subsection{Experimental Setup}

\subsubsection{Experimental Platforms}

We evaluate CommBench on three representative GPU clusters spanning both NVIDIA and AMD ecosystems as well as intra-node and inter-node communication. The first platform consists of a single node equipped with 8 NVIDIA B300 GPUs interconnected via NVLink. The second platform comprises NVIDIA GH200 nodes connected through 400\,Gb/s InfiniBand using ConnectX-7 NICs and BlueField-3 DPUs, enabling inter-node RDMA evaluation. The third platform consists of AMD Instinct MI325X nodes interconnected by Infinity Fabric (XGMI) within each node and 400\,Gb/s RoCE across nodes. Together, these platforms allow CommBench to evaluate GPU communication programs over realistic NVLink, InfiniBand, and RoCE communication paths on both CUDA and ROCm/HIP software stacks.

\subsubsection{Model Configuration}

We evaluate GPT-5.5, Gemini-3.1-Pro-Preview, Claude Opus 4.7, GLM-5.1, Kimi-K2.6, DeepSeek-V4-Pro, and Qwen3.7-Max~\cite{openai2026gpt55,google2026gemini31propreview,anthropic2026claudeopus47,zhipuai2026glm51,moonshotai2026kimik26,deepseekai2026deepseekv4,alibabacloud2026qwen37max}. Unless otherwise specified, all models are executed using their default reasoning modes and decoding parameters, including temperature, top-$p$, and other model-specific settings. GPT-5.5 and DeepSeek-V4-Pro use \texttt{medium} reasoning effort; Gemini-3.1-Pro-Preview uses \texttt{high}; Claude Opus 4.7 uses adaptive thinking; and GLM-5.1 and Kimi-K2.6 use their default thinking modes. Qwen3.7-Max is evaluated with thinking disabled to avoid excessive inference latency and timeout failures. No model-specific prompt engineering or hyperparameter tuning is applied.

% Unless otherwise specified, all evaluated models are executed using their default reasoning modes and default decoding parameters, including temperature, top-$p$, and other model-specific settings. Specifically, GPT-5.5 uses \texttt{medium} reasoning effort; Gemini-3.1-Pro-Preview uses \texttt{high}; Claude Opus 4.7 uses adaptive thinking; GLM-5.1 and Kimi-K2.6 enable their default thinking modes; DeepSeek-V4-Pro uses \texttt{medium}; and Qwen3.7-Max is evaluated with thinking disabled to avoid excessive inference latency and timeout failures. No model-specific prompt engineering or hyperparameter tuning is applied.

\subsection{Evaluation Metrics}

We evaluate generated programs from four complementary perspectives: correctness, performance, efficiency, and cost.

\paragraph{Pass Rate.}
Pass Rate measures the fraction of benchmark tasks whose generated implementations compile successfully, execute correctly, and pass all correctness checks.

\paragraph{PASS+Good.}
PASS+Good reports the fraction of benchmark tasks whose generated implementations are both correct and performant, achieving at least 95\% of the reference implementation's performance. Performance is evaluated using the primary efficiency metric appropriate for each task, such as latency or throughput.

\paragraph{GM-Speedup.}
For each passing benchmark task, we compute a speedup score by taking the geometric mean of the generated-to-reference performance ratio across all evaluated input sizes. For metrics where lower values are better (e.g., latency), we first convert them to their reciprocals so that higher values consistently indicate better performance. The overall GM-Speedup is the geometric mean of these per-task scores over all passing examples.

Using the geometric mean across input sizes ensures that each workload contributes equally regardless of its absolute throughput, preventing large workloads from dominating the aggregate metric.

\paragraph{Quality-Weighted Pass Rate.}
Because GM-Speedup is computed only over passing examples, it does not account for a model's overall success rate. Consequently, a model that solves more (and often more challenging) tasks may achieve a lower GM-Speedup despite exhibiting stronger overall capability. To jointly capture correctness coverage and generated code quality, we define the \emph{Quality-Weighted Pass Rate} as follows:

\[
\mathrm{Pass}\times\mathrm{GM}
=
\mathrm{PassRate}
\times
\mathrm{GM\text{-}Speedup}
\]

as our primary ranking metric, jointly capturing correctness coverage and generated code quality.

\paragraph{Cost.}
For proprietary models, we additionally report the average API cost (USD) per benchmark example.

\subsubsection{Performance Categories}

Following common benchmarking practice, we categorize the performance of each correct implementation into four levels according to its speed relative to the reference implementation:

\begin{itemize}
    \item \textbf{Better}: at least 20\% faster than the reference;
    \item \textbf{Comparable}: within $[-5\%,+20\%]$ of the reference;
    \item \textbf{Degraded}: between $-40\%$ and $-5\%$;
    \item \textbf{Severely Degraded}: more than 40\% slower than the reference.
\end{itemize}

\subsection{Results}
\paragraph{Overall Results.}
Table~\ref{tab:leaderboard} summarizes the overall performance of the evaluated models on CommBench. GPT-5.5 achieves the strongest results, with the highest Quality-Weighted Pass Rate of 0.467, Pass Rate of 57.4\%, and PASS+Good rate of 30.7\%, substantially outperforming the other models. Notably, the best-performing model does not necessarily achieve the highest GM-Speedup. A stronger model may solve a broader set of tasks, including more challenging ones, but with degraded performance on some of them, thereby lowering its average speedup. In contrast, a lower-ranked model may succeed only on relatively easy tasks while achieving performance comparable to the reference implementations, resulting in a higher GM-Speedup. Overall, current models can occasionally generate highly optimized GPU communication code, but they still struggle to do so reliably across diverse tasks.

\subsection{Analysis: Top vs. Bottom Models}

To better understand the sources of performance variation on CommBench, we conduct a detailed comparison between the highest- and lowest-scoring models, GPT-5.5 (Quality-Weighted Pass Rate 0.467) and DeepSeek-V4-Pro (Quality-Weighted Pass Rate 0.197). We analyze their behavior along four axes: difficulty level, task type, library coverage, and the runtime performance of the generated code.

\begin{figure}[t]
\centering
\begin{minipage}{0.49\linewidth}
\centering
\includegraphics[width=\linewidth]{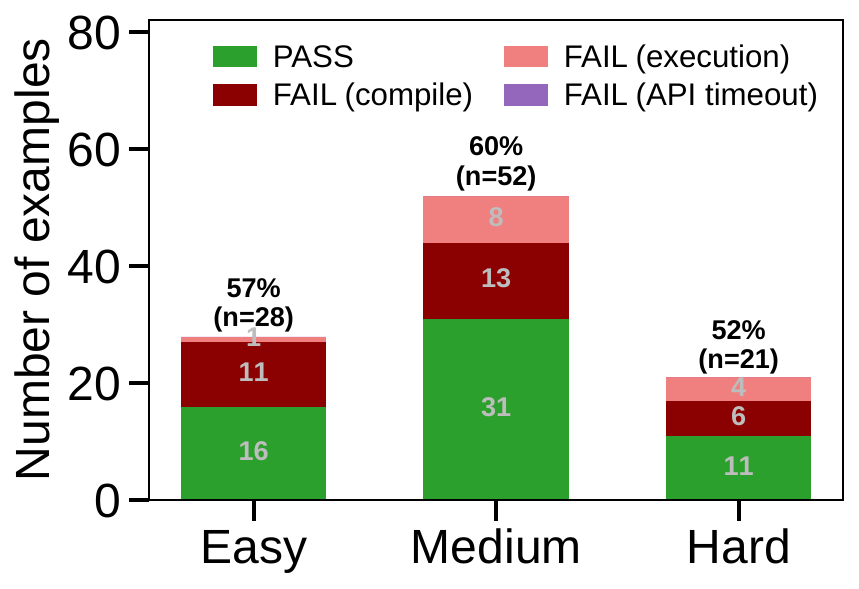}\\
{\small (a) GPT-5.5}
\end{minipage}
\hfill
\begin{minipage}{0.49\linewidth}
\centering
\includegraphics[width=\linewidth]{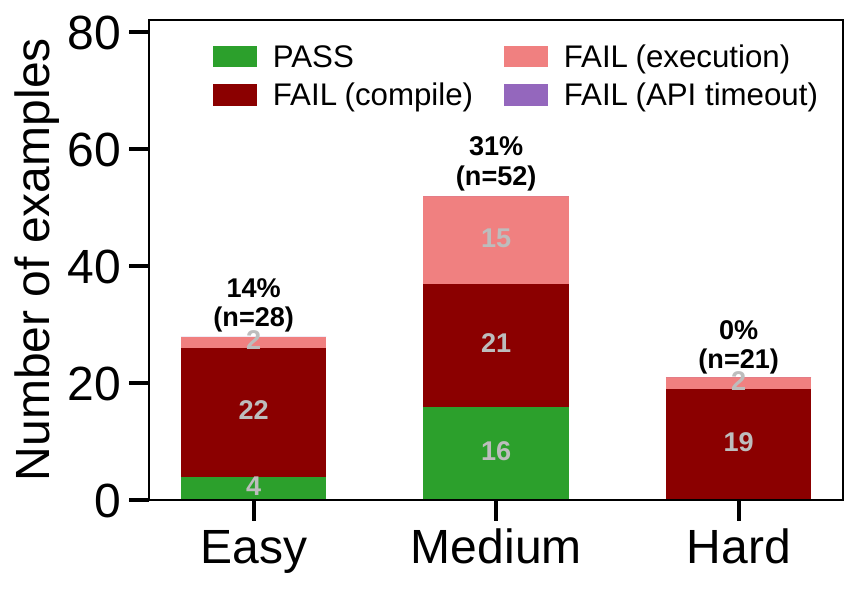}\\
{\small (b) DeepSeek-V4-Pro}
\end{minipage}
\caption{Pass rate by human-annotated difficulty level.}
\label{fig:difficulty}
\end{figure}

\subsubsection{Difficulty Breakdown.}
Human-annotated difficulty does not always coincide with the difficulty a model actually experiences (Figure~\ref{fig:difficulty}). DeepSeek-V4-Pro performs poorly even on Easy examples (14\%), a weakness largely attributable to gaps in library-specific API knowledge. For libraries such as MSCCL++ and ThunderKittens, the model lacks the interface semantics needed to invoke APIs correctly, and it therefore fails on examples that require only a handful of function calls to realize otherwise straightforward functionality. At the Hard end of the spectrum, many examples ask the model to implement communication primitives essentially from scratch (e.g., AllReduce over NVLink or RDMA), which exposes DeepSeek-V4-Pro's limited capability on complex GPU communication and computation tasks.

\begin{figure}[t]
\centering

\includegraphics[width=0.9\linewidth]{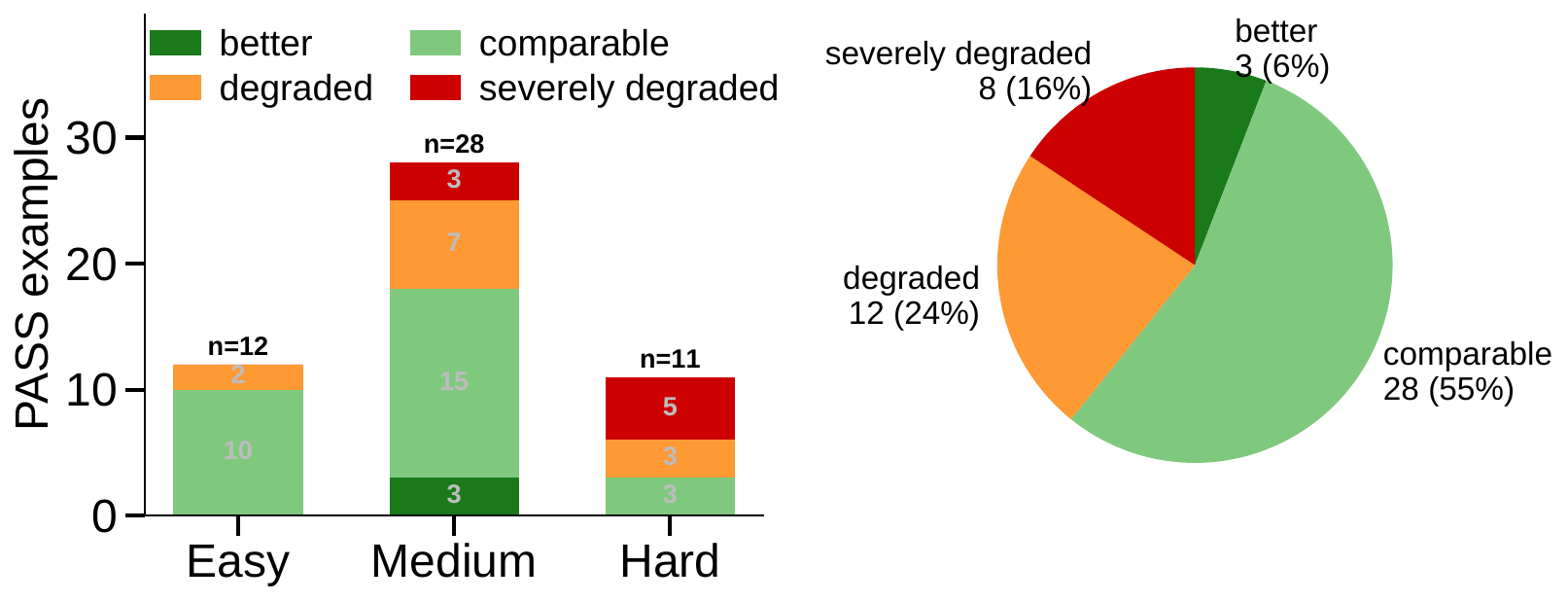}\\
{\small (a) GPT-5.5}

\vspace{0.5em}

\includegraphics[width=0.9\linewidth]{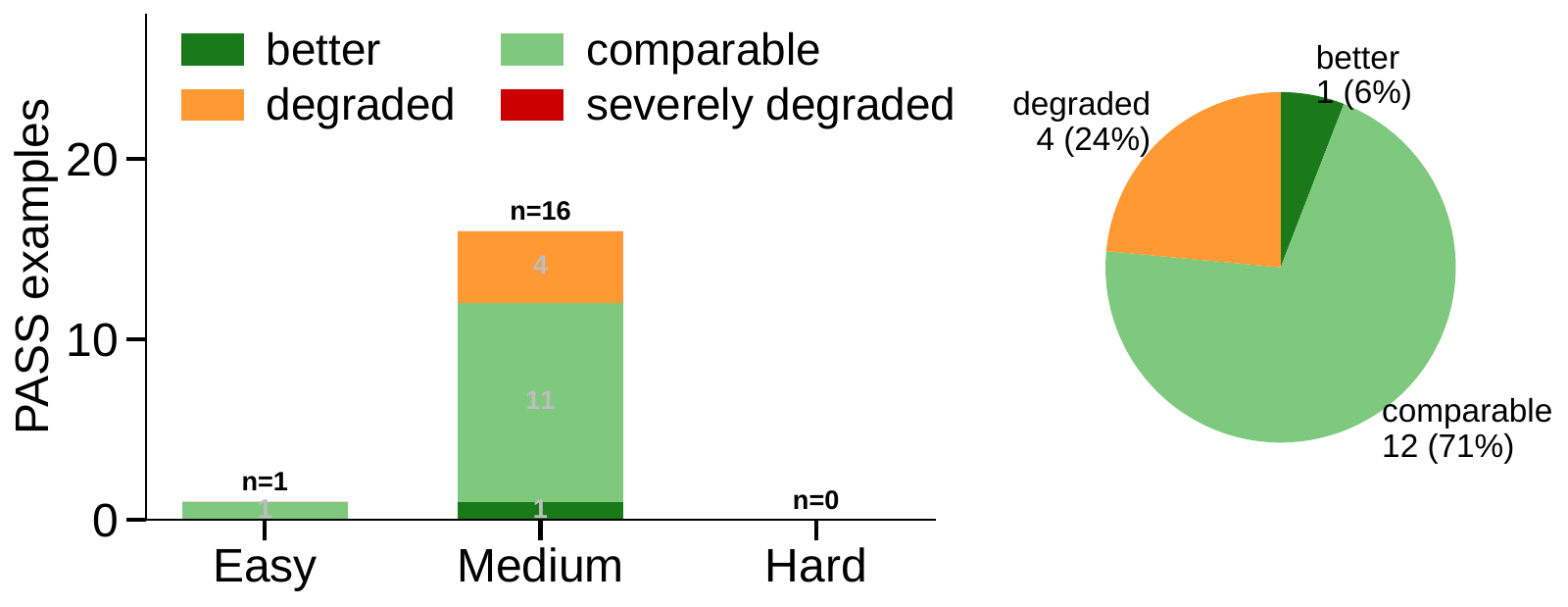}\\
{\small (b) DeepSeek-V4-Pro}

\caption{Distribution of performance categories over passing examples. We count only examples that report meaningful performance metrics.}
\label{fig:pass-quality}
\end{figure}

\begin{figure}[!t]
\centering

\includegraphics[width=0.9\linewidth]{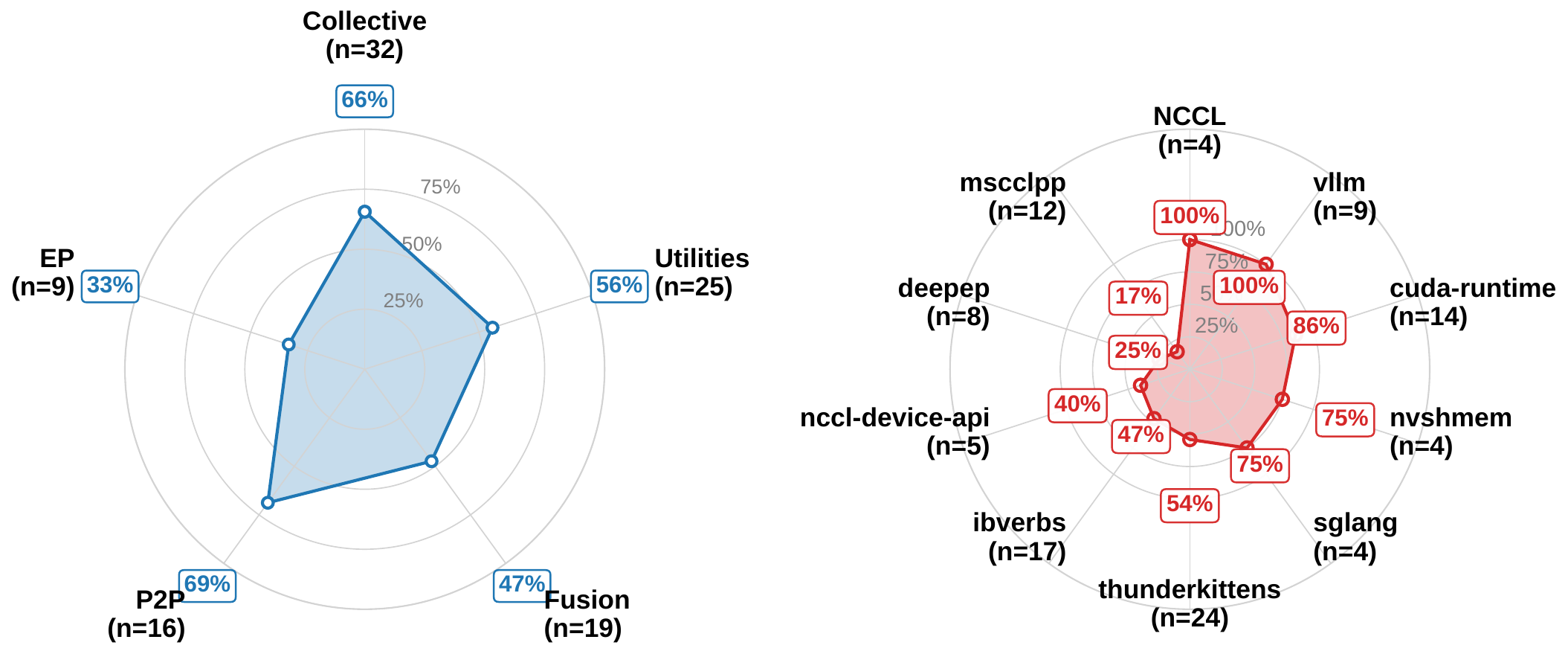}\\
{\small (a) GPT-5.5}

\vspace{0.5em}

\includegraphics[width=0.9\linewidth]{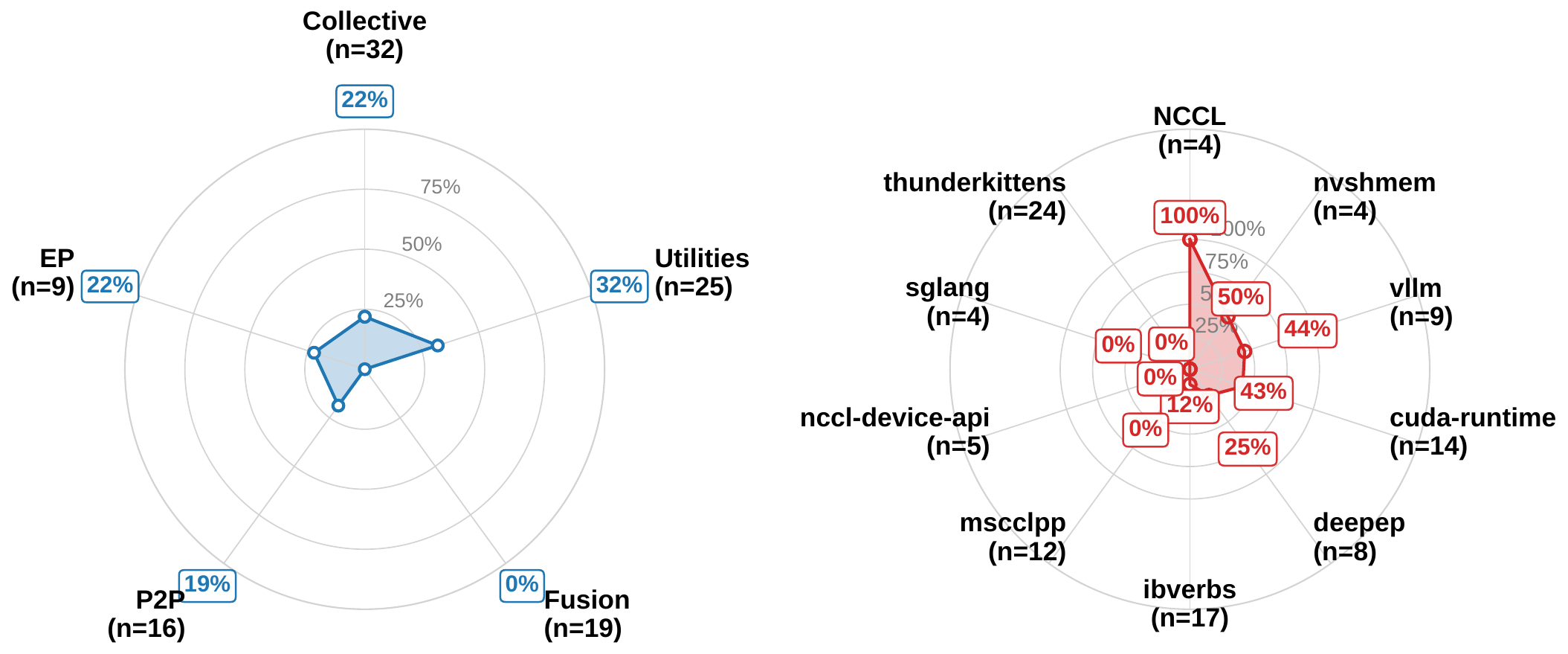}\\
{\small (b) DeepSeek-V4-Pro}

\caption{Per-tag and per-library pass rates.}
\label{fig:coverage}
\end{figure}

\subsubsection{Performance Quality Among Passing Examples.}
Among passing examples, GPT-5.5 exhibits the widest quality spread (Figure~\ref{fig:pass-quality}): eight severely degraded cases (16\% of the 51 passing examples with meaningful performance measurements), all concentrated in Medium and Hard examples. Because this distribution is computed only over passing examples, tasks that GPT-5.5 successfully compiles and executes, albeit with degraded performance, may simply fail to compile or run under DeepSeek-V4-Pro and therefore never appear in its performance distribution.

\subsubsection{Tag and Library Coverage.}
GPT-5.5 dominates on Collective tasks (66\% vs.\ 22\%) and is the only model with meaningful coverage of specialized libraries: MSCCL++ (17\%), the NCCL Device API (40\%), and ThunderKittens (54\%). DeepSeek-V4-Pro is competitive on NCCL (100\%) and on P2P tasks, but scores 0\% across all three specialized libraries (Figure~\ref{fig:coverage}). For widely adopted libraries such as vLLM, NCCL, and NVSHMEM, both models perform well.

\begin{table}[t]
\centering
\small
\caption{Cumulative passing examples for DeepSeek-V4-Pro across five self-correction rounds.}
\label{tab:cumulative}
\begin{tabular}{cccc}
\toprule
Round & Cumulative PASS & Pass Rate & New this round \\
\midrule
1 & 20 & 19.8\% & \\
2 & 28 & 27.7\% & $+8$ \\
3 & 32 & 31.7\% & $+4$ \\
4 & 34 & 33.7\% & $+2$ \\
5 & 42 & 41.6\% & $+8$ \\
\bottomrule
\end{tabular}
\end{table}

\begin{figure}[t]
\centering
\begin{minipage}{0.49\linewidth}
\centering
\includegraphics[width=\linewidth]{fig/fig1_level_breakdown_deepseek-v4-pro.pdf}\\
{\small (a) max rounds $=1$}
\end{minipage}
\hfill
\begin{minipage}{0.49\linewidth}
\centering
\includegraphics[width=\linewidth]{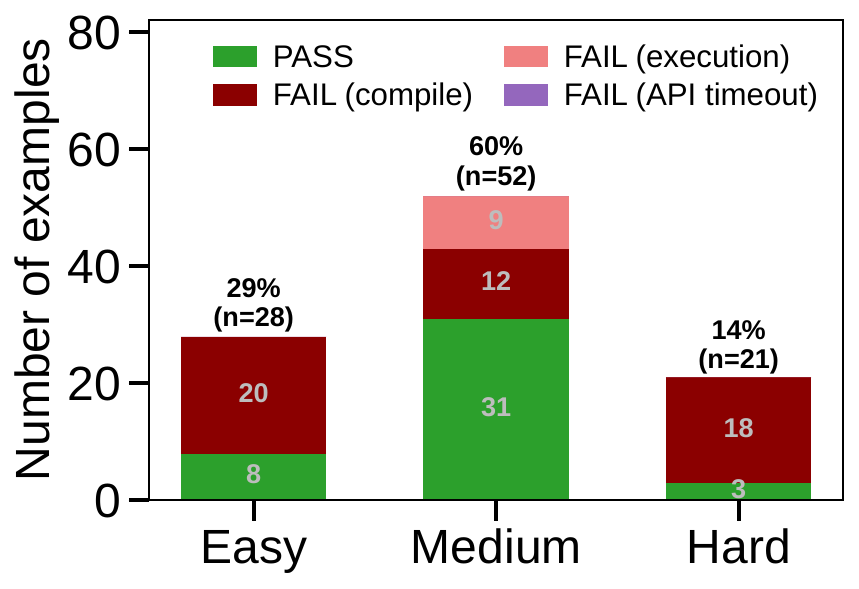}\\
{\small (b) max rounds $=5$}
\end{minipage}
\caption{DeepSeek-V4-Pro difficulty breakdown with a single round versus five refinement rounds.}
\label{fig:max5-difficulty}
\end{figure}

\begin{figure}[t]
  \centering

  \includegraphics[width=0.9\linewidth]
  {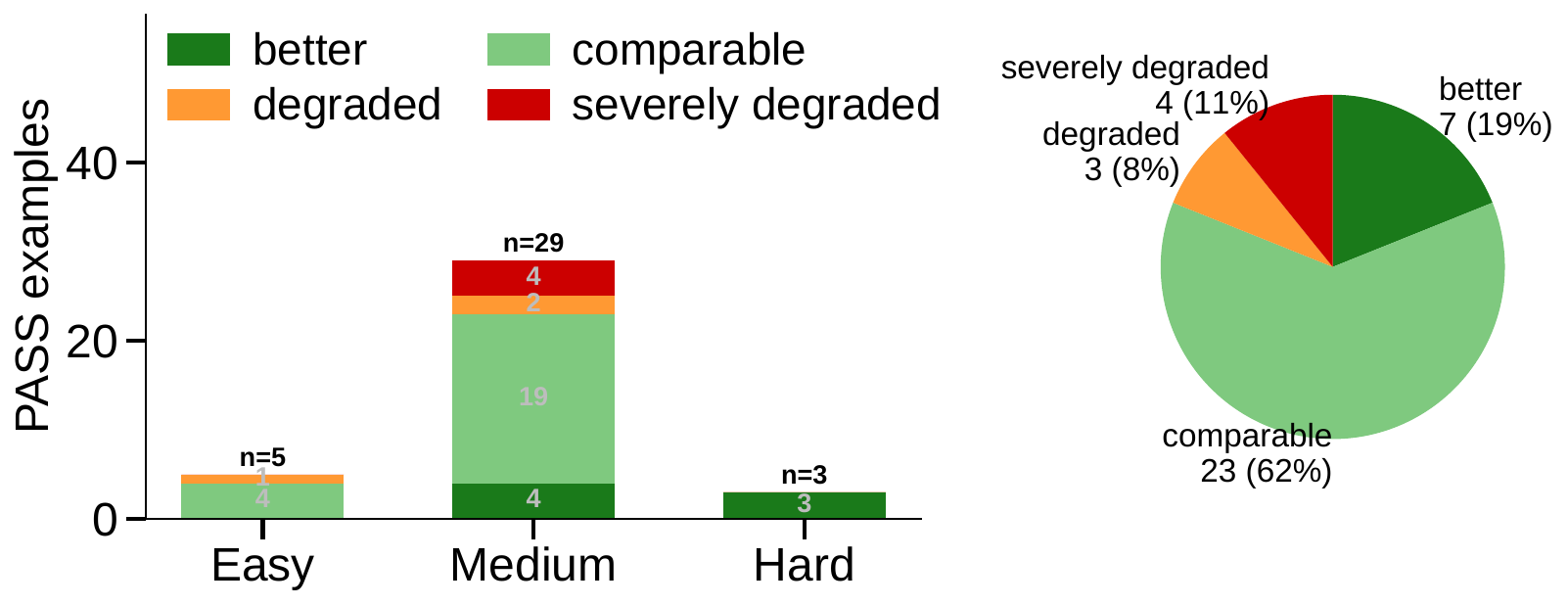}\\
  {\small (a) Performance among passing examples}

  \vspace{0.5em}

  \includegraphics[width=0.9\linewidth]
  {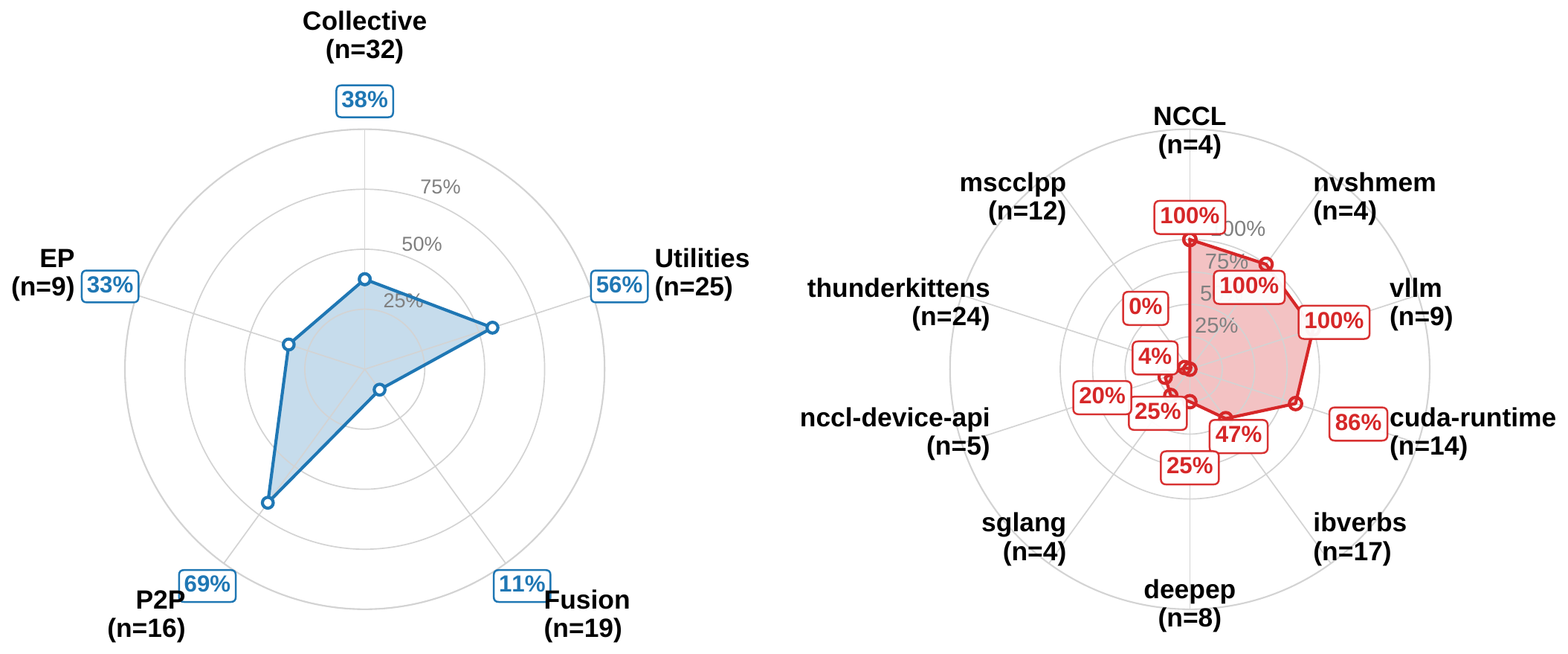}\\
  {\small (b) Tag and library coverage}

  \caption{DeepSeek-V4-Pro with five refinement rounds.}
  \label{fig:max5-quality}
\end{figure}

\subsubsection{Effect of Iterative Refinement.}
Due to budget constraints, GPT-5.5 is evaluated with a single generation round. Despite its weaker first-round performance, DeepSeek-V4-Pro costs only about 1\% as much per example (\$0.02 vs.\ \$1.91), making multi-round self-correction economically viable. As shown in Table~\ref{tab:cumulative} and Figure~\ref{fig:max5-difficulty}, five rounds more than double its pass rate (from 20 to 42 examples, or 19.8\% to 41.6\%) and improve its pass rate on Medium-difficulty commodity-library tasks from 31\% to 60\%. However, they do not unlock Hard examples or specialized libraries (MSCCL++: 0\%; ThunderKittens: 4\%), which require missing domain knowledge. Thus, DeepSeek-V4-Pro with retries is reasonable when tasks use commodity libraries (NCCL, vLLM, CUDA Runtime, and NVSHMEM) and a retry budget is acceptable.

\begin{table}[!t]
\centering
\footnotesize
\setlength{\tabcolsep}{3.5pt}
\renewcommand{\arraystretch}{1.1}
\caption{Cost-matched comparison of the three strongest models.}
\label{tab:cost-matched}
\begin{tabular}{lcccc}
\toprule
Model &
\makecell{Quality-\\Weighted\\Pass Rate $\uparrow$} &
\makecell{Pass\\Rate\\(\%) $\uparrow$} &
\makecell{PASS+\\Good\\(\%) $\uparrow$} &
\makecell{GM-\\Speedup\\$\uparrow$} \\
\midrule
Claude Opus 4.7        & \textbf{0.690} & 61.4          & 37.6          & \textbf{1.125} \\
Gemini-3.1-Pro-Preview & 0.689          & \textbf{66.3} & \textbf{45.5} & 1.039          \\
GPT-5.5                & 0.467          & 57.4          & 30.7          & 0.813          \\
\bottomrule
\end{tabular}
\end{table}

\subsubsection{Cost-Matched Comparison.}
We select the
three strongest models and repeat the
comparison under an approximately equal \emph{dollar} budget: the two cheaper models, Claude Opus 4.7
and Gemini-3.1-Pro-Preview, are permitted substantially more rounds in which to
correct and tune their code, whereas GPT-5.5 is restricted to a single round.As in Table~\ref{tab:cost-matched}, once
spend rather than rounds is held fixed, both cheaper models overtake GPT-5.5 by
a wide margin on the Quality-Weighted Pass Rate ($0.690$ and $0.689$ versus
$0.467$), and the gap is driven by both of its factors: they pass more examples
(61.4\% and 66.3\% versus 57.4\%) \emph{and} the code they eventually produce is
faster (GM-Speedup $1.125$ and $1.039$ versus $0.813$). Iterative refinement
thus buys more than correctness. Rounds that begin as compile-and-crash repairs
end as performance tuning, so the additional budget also lifts the quality of
the code that already passed.

\subsection{Case Studies}

We now examine three representative examples, one from each difficulty level, that illustrate distinct failure modes: an Easy task passed by the only two models, a Medium task failed by every model despite requiring only straightforward API usage, and a Hard task failed by every model because it combines a niche library with minimal algorithmic guidance.

\subsubsection{Case 1: Partial Pass on ThunderKittens AllToAll.}
This example (Easy; Collective; \texttt{thunderkittens}) asks the model to implement a BF16 multi-GPU AllToAll kernel using the ThunderKittens library, employing the Tensor Memory Accelerator (TMA) for tile-level data movement between GPU shards over NVLink. Only two small device-side kernels must be written: \texttt{all\_to\_all::kernel} (TMA-load one tile from the local input, then TMA-store it to the destination device's output shard) and \texttt{all\_to\_all\_barrier::kernel} (a one-line \texttt{barrier\_all} synchronization). All host scaffolding is provided.

Both kernels are under ten lines and require only calls to existing ThunderKittens interfaces. The root cause of failure is therefore ThunderKittens-specific API knowledge: correct type scoping, the correct \texttt{tma::load\_async} argument signature, and the \texttt{shared\_allocator::allocate} template usage. ThunderKittens is a niche research library with minimal training-data coverage, which leads models to mis-scope types, hallucinate non-existent APIs, or supply wrong argument forms (Table~\ref{tab:case1}).

\begin{table*}[t]
\centering
\footnotesize
\setlength{\tabcolsep}{4pt}
\renewcommand{\arraystretch}{1.12}
\caption{Per-model outcomes on the ThunderKittens AllToAll example (Case 1).}
\label{tab:case1}

\begin{tabularx}{\textwidth}{
    @{}
    >{\raggedright\arraybackslash}p{0.16\textwidth}
    >{\centering\arraybackslash}p{0.06\textwidth}
    >{\centering\arraybackslash}p{0.13\textwidth}
    >{\centering\arraybackslash}p{0.16\textwidth}
    Y
    @{}
}
\toprule
Model & Rounds & Outcome & Perf.\ vs.\ Ref. & Failure Pattern \\
\midrule

GPT-5.5
& 1
& PASS
& Comparable ($-0.23\%$)
& --- \\

Claude Opus 4.7
& 1
& PASS
& Comparable ($+0.17\%$)
& --- \\

Gemini-3.1-Pro-Preview
& 1
& Compile error
& ---
& Used \texttt{shared\_tile} without the required namespace qualifier. \\

Kimi-K2.6
& 1
& Compile error
& ---
& Used the same unqualified \texttt{shared\_tile} reference. \\

GLM-5.1
& 1
& Compile error
& ---
& Passed an incompatible type to
\texttt{shared\_allocator::allocate}. \\

Qwen3.7-Max
& 1
& Compile error
& ---
& Passed incompatible argument types to
\texttt{tma::load\_async}. \\

\midrule

\multirow{2}{*}{DeepSeek-V4-Pro}
& \multirow{2}{*}{5}
& \multirow{2}{*}{Compile/link error}
& \multirow{2}{*}{---}
& \textbf{Round 1:} Hallucinated unsupported ThunderKittens
synchronization and TMA APIs. \\

&
&
&
&
\textbf{Rounds 2--5:} Repeatedly produced invalid CUDA/C++ code due
to incorrect program structure, API usage, and type handling. \\

\bottomrule
\end{tabularx}
\end{table*}

\begin{table*}[!t]
\centering
\footnotesize
\setlength{\tabcolsep}{4pt}
\renewcommand{\arraystretch}{1.12}
\caption{Per-model outcomes on the GPU Barrier Within a CTA example (Case 2).}
\label{tab:case2}

\begin{tabularx}{\textwidth}{
    @{}
    >{\raggedright\arraybackslash}p{0.18\textwidth}
    >{\centering\arraybackslash}p{0.07\textwidth}
    >{\centering\arraybackslash}p{0.15\textwidth}
    Y
    @{}
}
\toprule
Model & Rounds & Outcome & Failure Pattern \\
\midrule

GPT-5.5
& 1
& Deadlock
& The consumer executes before the producers because of warp branch
ordering; without a \texttt{kProbeBound} guard, the kernel hangs
indefinitely. \\

Claude Opus 4.7
& 1
& Correctness FAIL
& The consumer executes before the producers and incorrectly applies
\texttt{phase \^{}= 1}, although \texttt{mbarrier.init} resets the parity
to zero in each iteration. \\

Gemini-3.1-Pro-Preview
& 1
& Correctness FAIL
& The consumer executes before the producers, exits through
\texttt{kProbeBound}, and reads an uninitialized tile. \\

Kimi-K2.6
& 1
& Compile error
& Hallucinated a nonexistent CUDA intrinsic. \\

GLM-5.1
& 1
& Runtime error
& Used inconsistent inline-assembly operand indices. \\

Qwen3.7-Max
& 1
& Correctness FAIL
& Used the wrong PTX register constraint, specifying \texttt{"l"}
instead of \texttt{"r"}. \\

\midrule

\multirow{5}{*}{DeepSeek-V4-Pro}
& \multirow{5}{*}{5}
& \multirow{5}{*}{Compile error}
& \textbf{Round 1:} Used an incorrect PTX register constraint. \\

&
&
& \textbf{Round 2:} Omitted the destination operand required by
\texttt{mbarrier.arrive}. \\

&
&
& \textbf{Round 3:} Retained the same incomplete
\texttt{mbarrier.arrive} instruction. \\

&
&
& \textbf{Round 4:} Failed to correct the missing destination operand
despite compiler feedback. \\

&
&
& \textbf{Round 5:} Repeated the same invalid PTX instruction and
remained unable to compile. \\

\bottomrule
\end{tabularx}
\end{table*}

\begin{table*}[!t]
\centering
\footnotesize
\setlength{\tabcolsep}{4pt}
\renewcommand{\arraystretch}{1.12}
\caption{Per-model outcomes on the MSCCL++ AllToAll example (Case 3).}
\label{tab:case3}

\begin{tabularx}{\textwidth}{
    @{}
    >{\raggedright\arraybackslash}p{0.18\textwidth}
    >{\centering\arraybackslash}p{0.07\textwidth}
    >{\centering\arraybackslash}p{0.14\textwidth}
    Y
    @{}
}
\toprule
Model & Rounds & Outcome & Failure Pattern \\
\midrule

GPT-5.5
& 1
& Compile error
& Included several MSCCL++ headers but omitted the header defining
\texttt{MemoryChannel}; it also referenced an incorrect class name. \\

Claude Opus 4.7
& 1
& Compile error
& Hallucinated unsupported MSCCL++ APIs. \\

Gemini-3.1-Pro-Preview
& 1
& Compile error
& Hallucinated unsupported MSCCL++ APIs. \\

Kimi-K2.6
& 1
& Compile error
& Hallucinated unsupported MSCCL++ APIs. \\

GLM-5.1
& 1
& Compile error
& Hallucinated unsupported MSCCL++ APIs. \\

Qwen3.7-Max
& 1
& Compile error
& Hallucinated unsupported MSCCL++ APIs. \\

\midrule

\multirow{2}{*}{DeepSeek-V4-Pro}
& \multirow{2}{*}{5}
& \multirow{2}{*}{Compile error}
& \textbf{Round 1:} Used MSCCL++ types without including the required
headers, leaving the namespace undefined. \\

&
&
& \textbf{Rounds 2--5:} Invented a different nonexistent MSCCL++ header
in each refinement round. \\

\bottomrule
\end{tabularx}
\end{table*}

\subsubsection{Case 2: A Task Every Model Failed.}
This example (Medium; Utilities; \texttt{cuda-runtime}) asks the model to implement intra-CTA producer/consumer synchronization using a Hopper/Blackwell shared-memory \texttt{mbarrier} with a non-blocking \texttt{try\_wait} probe, a pattern used in persistent-kernel tile pipelines (e.g., CUTLASS and Mirage) to overlap data arrival with computation without stalling the warp scheduler. The model must implement three inline-PTX device helpers (\texttt{initialize\_barrier}, \texttt{arrive}, \texttt{try\_wait\_barrier}) and a benchmark kernel body.

This task requires only the use of a small set of well-defined \texttt{mbarrier} interfaces, which is exactly the kind of API memorization at which LLMs are expected to excel. However, \texttt{mbarrier} is Hopper-specific PTX introduced in \texttt{sm\_90}, and models appear to lack sufficient training data for these instructions, making them prone to mis-remembering argument counts, ordering, or address-space requirements (Table~\ref{tab:case2}).

\subsubsection{Case 3: A Niche-Library API with No Algorithmic Hints.}
This example (Hard; Collective; \texttt{mscclpp}) asks the model to implement the fastest intra-node AllToAll kernel using MSCCL++ \texttt{MemoryChannel} primitives. The template provides no algorithmic description, only a minimal comment (``implement the fastest All to All CUDA kernel'') and a set of empty \texttt{// TODO} stubs. The model must produce one GPU kernel (\texttt{alltoall2}) that uses \texttt{MemoryChannel} for direct peer writes, together with the full host-side \texttt{All2All} class (constructor, buffer allocation, channel setup, launch, correctness verification, and barrier).

The task thus requires implementing a functionally complete, semantically correct intra-node AllToAll from scratch, all consistent with the MSCCL++ \texttt{MemoryChannel} interface contract and with no algorithmic hints. This is difficult even for an expert programmer unfamiliar with MSCCL++. Every model failed at compile time: MSCCL++ headers and class interfaces are largely absent from training data, causing models to hallucinate non-existent header paths and method signatures before reaching any algorithmic logic (Table~\ref{tab:case3}).

\section{Related Work}

\subsubsection{LLMs for GPU and parallel code generation.}
LLMs have shown substantial promise in code generation, program repair, and performance optimization, motivating growing efforts to improve their ability to generate efficient GPU programs. Earlier learning-based approaches such as BabelTower~\cite{pmlr-v162-wen22b} studied the translation of sequential programs into CUDA. More recent systems improve GPU kernel generation and optimization through domain-specific post-training~\cite{kernelllm2025,kong2025concurconcisenessmakesstateoftheart,li2025autotritonautomatictritonprogramming,woo2026tritonrltrainingllmsthink}, agentic execution and profiling feedback~\cite{wei2025astramultiagentgpukernel,dong2025starkstrategicteamagents,du2025akgkernelagentmultiagent,sun2026kernelskillmultiagentframeworkgpu,jaber2026autokernelautonomousgpukernel,gai2026optimizingcudalikehuman}, evolutionary and structured search~\cite{ran2026kernelbandsteeringllmbasedkernel,cao2026ksearchllmkernelgeneration,wiedemann2026kernelfoundryhardwareawareevolutionarygpu,du2026kernelsmithunifiedrecipeevolutionary}, or reinforcement learning~\cite{baronio2025kevinmultiturnrlgenerating,li2026cudal1improvingcudaoptimization,su2026cudal2surpassingcublasperformance,liu2026drkernelreinforcementlearning,dai2026cudaagentlargescaleagentic,10.1609/aaai.v40i34.40155}. Most of these systems focus on single-GPU computation. Broader efforts target MPI code generation~\cite{schneider2024mpirigenmpicodegeneration,10.1145/3773656.3773659}, parallel computing~\cite{kaplan2026paracodexprofilingguidedautonomouscoding}, and general HPC code optimization~\cite{kadosh2024monocoderdomainspecificcodelanguage,chaturvedi2024hpccoderv2studyingcodellms,hayashi2026vibecodehpcagentbasediterativeprompting}. CUCo~\cite{varadharajan2026cucoagenticframeworkcompute} is more directly related to multi-GPU communication, using an agentic workflow to transform host-driven CUDA and NCCL programs into compute–communication co-designed kernels. 

% However, GPU communication programming also requires reasoning about cross-rank coordination, communication APIs, topology and network setup, memory registration, and host–device synchronization. Dedicated benchmarks are therefore needed to evaluate and advance LLM capabilities for GPU communication code generation.

\subsubsection{Benchmarks for LLM-Generated GPU and parallel code.}
Existing benchmarks evaluate LLM-generated parallel and GPU programs at several levels. ParEval~\cite{10.1145/3625549.3658689} covers scientific and parallel programming across multiple programming models, while KernelBench~\cite{ouyang2025kernelbenchllmswriteefficient} provides a widely adopted evaluation of functionally correct and performant GPU kernels generated from PyTorch workloads. Subsequent benchmarks extend this setting along several dimensions: ComputeEval~\cite{computeeval}, robust-kbench~\cite{lange2025robustagenticcudakernel}, and CUDAEval~\cite{gong2025largesmalltransferringcuda} broaden evaluation beyond PyTorch operators to more general CUDA programming tasks; TritonBench~\cite{li2025tritonbenchbenchmarkinglargelanguage} and KernelBenchX~\cite{wang2026kernelbenchxcomprehensivebenchmarkevaluating} focus on Triton generation; MultiKernelBench~\cite{wen2025multikernelbenchmultiplatformbenchmarkkernel} extends evaluation across heterogeneous accelerators; FlashInfer-Bench~\cite{xing2026flashinferbenchbuildingvirtuouscycle} and Atrex-Bench~\cite{yang2026llmgeneratedgpukernelsproductionready} incorporate serving traces and production-oriented inference workloads; while CUDABench~\cite{zhu2026cudabenchbenchmarkingllmstexttocuda} and SOL-ExecBench~\cite{lin2026solexecbenchspeedoflightbenchmarkingrealworld} assess performance relative to hardware-aware efficiency targets. Despite these extensions, most benchmarks predominantly focus on single-device computation. While ParallelKernelBench~\cite{chan-parallelkernelbench-2026}, the closest concurrent work, focuses on rewriting PyTorch programs with NCCL operations into optimized intra-node NVLink kernels, CommBench spans both intra- and inter-node communication across broader workloads, including collectives, MoE expert-parallel communication, and communication--computation fusion. Existing benchmarks provide limited coverage of GPU communication code across diverse programming abstractions, inter-node communication via RoCE and InfiniBand, and cross-vendor accelerator ecosystems spanning NVIDIA and AMD.

\section{Discussion}

\paragraph{Post-training for knowledge-related failures.}
Many CommBench failures are knowledge-related: models hallucinate headers, types, and method signatures, even on Easy tasks requiring few API calls. Coverage drops sharply on specialized libraries, with most models solving few or no tasks from MSCCL++, ThunderKittens, and the NCCL Device API. Multi-round correction offers limited relief because compiler feedback can repair familiar code but cannot provide unseen interface semantics; improvements therefore concentrate on common libraries and easier tasks. A promising remedy is targeted post-training on CommBench-style data. Supervised fine-tuning on expert-written communication code can teach correct headers, type scoping, and signatures underrepresented in web corpora, while reinforcement learning against the executable harness can reward both correctness and efficiency. Complementarily, retrieval augmentation can supply documentation, headers, and API signatures at inference time, grounding generation in the actual interface contract.

\bibliography{aaai2027}


\end{document}